\documentclass[aps,prl,twocolumn,superscriptaddress,showpacs,
floatfix,nofootinbib]{revtex4-1}

\usepackage{graphicx}
\usepackage{amsmath}
\usepackage{color}

\begin{document}
\title{From the lightest nuclei to the equation of state of asymmetric nuclear matter
with realistic nuclear interactions}

\author{S. Gandolfi}
\affiliation{Theoretical Division, Los Alamos National Laboratory, Los Alamos, NM, 87545}
\author{A. Lovato}
\affiliation{Argonne Leadership Computing Facility, Argonne National Laboratory, Argonne, IL 60439}
\affiliation{Physics Division, Argonne National Laboratory, Argonne, IL 60439}
\author{J. Carlson}
\affiliation{Theoretical Division, Los Alamos National Laboratory, Los Alamos, NM, 87545}
\author{Kevin E. Schmidt}
\affiliation{Department of Physics, Arizona State University, Tempe, AZ 85287, USA}

\begin{abstract}
We present microscopic calculations of light and medium mass nuclei and
the equation of state of symmetric and asymmetric nuclear matter using
different nucleon-nucleon forces, including a new Argonne version
that has the same spin/isospin structure as local chiral forces at
next-to-next-to-leading order (N2LO). The calculations are performed
using Auxiliary Field Diffusion Monte Carlo (AFDMC) combined with an
improved variational wave function. We show that the AFDMC method can now
be used to successfully calculate the energies of very light to medium
mass nuclei as well as the energy of isospin-asymmetric nuclear matter,
demonstrating microscopically the quadratic dependence of the energy on
the symmetry energy.
\end{abstract}
\date{\today}


\maketitle

The knowledge of the equation of state (EOS) of nuclear matter, particularly
with arbitrary isospin asymmetry, i.e., different proton and neutron fractions, is
of fundamental importance for both nuclear physics and astrophysics.
The saturation density and the energy per particle of nuclear
matter can be used to test properties of finite nuclear systems extrapolated 
to the thermodynamic limit. Moreover, the study of the EOS of asymmetric matter 
allows for the understanding of the behavior of the isospin asymmetry energy,
i.e., the ``symmetry energy'', and for constraining the bulk properties of nuclear 
density functionals, which are often used to predict the properties of heavy nuclei 
and nuclei with large neutron excesses. Experimentally accessing the properties 
of such nuclei constitutes a formidable task; however, they are needed to understand
the observed abundances of the elements, their production in r-process 
nucleosynthesis  in supernovae, and the properties of exotic neutron-rich nuclei 
found in the crust of neutron stars.

The calculation of the EOS of nuclear matter is one of the most
challenging problems for many-body nuclear physics, and to
date no completely satisfactory solution is available. The main reason
is the non-perturbative nature of realistic nuclear forces, even when
soft nucleon-nucleon interactions are employed.  The repulsive core and
the strong tensor components induce strong many-body correlations.
Because of strong spin-isospin dependent
correlations and the large number of particles, the calculation of
asymmetric nuclear matter properties is more difficult than for pure
neutron matter or for light nuclei.

The latest generation of modern nucleon-nucleon (NN) forces can
fit scattering data with very high precision, with $\chi^2\sim 1$
per data point. Accurate nuclear NN potentials include Argonne
AV18~\cite{Wiringa:1995}, CD-Bonn~\cite{Machleidt:2001}, and several
forms of chiral forces derived within the chiral effective field theory
(see for example~\cite{Entem:2003}).  The
 NN interactions are typically combined with three-body forces, in such
 a way that
the different nuclear Hamiltonians describe very accurately
properties of light nuclei~\cite{Pieper:2001,Maris:2013}, medium
nuclei~\cite{Hagen:2013,Binder:2013}, and homogeneous neutron
matter~\cite{Gandolfi:2009,Gandolfi:2012,Hebeler:2010}.

Several many-body methods have been developed to accurately
solve for the ground-state of light nuclei with interactions fit
to the NN scattering data. These include the Green's Function
Monte Carlo (GFMC)~\cite{Carlson:1987}, methods based on basis
expansions, i.e. No Core Shell Model~\cite{Navratil:2000},
No Core Full Configuration~\cite{Maris:2009}, Hyperspherical
Harmonics~\cite{Barnea:1999}, and others similar.  The coupled
cluster~\cite{Dean:2004}, the Self Consistent Green's
Function~\cite{Dickhoff:2004} (SCGF), and the in-medium
SRG~\cite{Tsukiyama:2011} methods are useful to study
medium nuclei.  Other approaches are based on performing unitary
transformation of the nuclear Hamiltonian with the goal of softening
the nuclear interactions and have a fast convergence using perturbation
theory~\cite{Hebeler:2011}. Recently, coupled cluster methods have been
extended to study nuclear matter~\cite{Baardsen:2013,Hagen:2014}.

Quantum Monte Carlo methods, such as GFMC and the Auxiliary
Field Diffusion Monte Carlo~\cite{Schmidt:1999} (AFDMC), have
proved to be accurate for predicting properties of nuclei up to
A=12~\cite{Pieper:2005,Lovato:2013,Lovato:2014} and neutron
matter~\cite{Gandolfi:2009,Gandolfi:2012}. Recently, new
local versions of chiral forces have been fitted to scattering
data, and can be included in GFMC and AFDMC. They have been
employed to study pure neutron matter~\cite{Gezerlis:2013}
and light nuclei with A=3,4~\cite{Lynn:2014}. The AFDMC method
has been also employed for studying nuclear matter and medium
nuclei~\cite{Gandolfi:2007,Gandolfi:2007b}, but the accuracy of these
calculations was limited by the poor variational wave functions and
numerical issues arising from time-step errors. In this work we show
that the limitations associated with both these issues can be overcome,
making the accuracy of AFDMC comparable to GFMC.

We present calculations of the EOS of symmetric and asymmetric nuclear
matter using modern NN forces.  Because most previous methods have
only calculated symmetric nuclear matter and/or neutron matter EOS,
often a quadratic dependence of the energy on the isospin asymmetry has
been assumed.
Studies of asymmetric nuclear matter at zero temperature have 
been performed only within variational Fermi Hypernetted Chain/Single
Operator Chain technique~\cite{Lagaris:1981b}, 
Bruekner-Hartree-Fock~\cite{Zuo:1999} theory or by means of perturbative 
approaches~\cite{Drischler:2014}. At finite temperature, SCGF has also 
been employed~\cite{Frick:2005}.

In this work we present Quantum Monte Carlo calculations of asymmetric
nuclear matter at saturation density, and demonstrate the quadratic
dependence of the isospin asymmetry energy. We also show the application
of AFDMC to light and medium-mass nuclei, including $^{16}$O and
$^{40}$Ca, and discuss the extension to open-shell nuclei, with the
inclusion of BCS-like\cite{Bardeen:1957} correlations. We will finally
introduce a technique suitable to include  three-body forces in future
AFDMC calculations, that can in principle be included in other methods
as well.

Quantum Monte Carlo simulations extract the ground state properties of 
a many-body system through the evolution in imaginary time $\tau$ of a 
trial wave function $\Psi_T$:
\begin{equation}
\Psi(\tau)=\exp[-(H-E_T)\tau]\Psi_T \,
\end{equation}
where $E_T$ is a parameter that controls the normalization of the wave function, and $H$ is the 
Hamiltonian of the system
\begin{equation}
H=\sum_{i=1}^A\frac{p_i^2}{2m}+\sum_{i<j}v_{ij} \,,
\end{equation}
and $v_{ij}$ a two-body NN potential.
In the limit of $\tau\rightarrow\infty$ the wave function $\Psi(\tau)$
converges to the lowest energy state not orthogonal to $\Psi_T$.
AFDMC calculations use the trial wave function $\Psi_T$ to minimize the
variance of the calculation and as a  constraint to control the fermion
sign problem~\cite{Zhang:2003}. In previous AFDMC calculations highly
simplified wave functions, without any tensor or other spin-isospin
dependent correlations, have been used.  Such trial wave functions are
not fully adequate to treat systems with both neutrons and protons. The
reason is that the tensor interaction in the $np$ ($T=0$) channel is
very large. It can be shown that the expectation value of the tensor
component of the NN potential is nearly zero if the tensor correlations
are not included in the variational wave function. Therefore, those
correlations are an essential feature of the nuclear wave function.

In addition, in most of the previous AFDMC calculations the
auxiliary fields were sampled using the method described in
Ref.~\cite{Zhang:2003}. We have found much better time-step
dependence by adopting the sampling technique typically used in GFMC
calculations. Within this technique, the auxiliary fields are sampled
from real gaussians, and a walker
is propagated according to the auxiliary fields having opposite signs
(we reverse the spatial moves and the spin-isospin rotations
separately), and the final walker is sampled from these two or four
choices according to their importance sampled  weight. This method
removes any time-step errors associated with gradients of the trial
wave function.

In this work we have considered the Argonne AV6$'$
interaction~\cite{Wiringa:2002}, and a new interaction that
we call AV7$'$ with an additional spin-orbit term added to AV6$'$
to improve the phase shift fit~\cite{supplemental}. This
interaction is identical to AV8$'$ in pure neutron systems, and is
adjusted to give the best reproduction of AV8$'$ in the $^3S_1 -
^3D_1$ coupled channels.  The extension of AFDMC to deal with this
isospin-independent spin-orbit is possible without any further
approximation~\cite{Sarsa:2003}. (The isospin-dependent spin-orbit
either would need to be included perturbatively, or by more sophisticated sampling
methods not described here).  The AV7$'$ force gives a much
better fit to the lower partial wave nucleon-nucleon  phase shifts than
AV6$'$. In addition, the spin/isospin  structure of AV7$'$ is the same
as local chiral forces up to next-to-next-to-leading order
(N$^2$LO), so that AFDMC can be easily extended to use chiral potentials
of the form of Ref.~\cite{Gezerlis:2013,Gezerlis:2014}.

In order to improve the algorithm and avoid the aforementioned limitations
associated with the simple trial wave functions, we have implemented a
trial wave function including tensor correlations:
\begin{equation}
\label{eq.wf}
\langle R,S|\Psi_T\rangle = \langle RS|
\left[\prod_{i<j}f_c(r_{ij})\right] \left [1+\sum_{i<j,p} f_p(r_{ij})O^p_{ij}
\right ]|\Phi\rangle
\end{equation}
where the $p$ sum is over the operators
$\boldsymbol\tau_i\cdot\boldsymbol\tau_j$,
$\boldsymbol\sigma_i\cdot\boldsymbol\sigma_j
\boldsymbol\tau_i\cdot\boldsymbol\tau_j$, and
$(3\boldsymbol{\sigma}_i\cdot\hat{r}_{ij}
\boldsymbol{\sigma}_j\cdot\hat{r}_{ij}
-\boldsymbol\sigma_i\cdot\boldsymbol\sigma_j)
\boldsymbol\tau_i\cdot\boldsymbol\tau_j$. 
This wave function is not extensive and not as accurate as the
one used in GFMC for light nuclei~\cite{Pudliner:1997}, but it has
substantial overlap with the tensor components, unlike the simple wave
functions used in previous AFDMC calculations. The major drawback of
the GFMC wave functions is that their evaluation requires a number of
operations exponentially growing with $A$. Instead, the evaluation
of the wave function of Eq.~(\ref{eq.wf}) requires order $A^3$
operations so that its calculation is feasible even for large
systems. The radial functions $f_p(r)$ are obtained by minimizing 
the two-body cluster contribution to the energy per particle of symmetric
nuclear matter at saturation density, as described
in Ref. \cite{Lagaris:1981}. All the variational parameters are determined by
minimizing the variational energy of a given nucleus, following the procedure 
described in Ref.~\cite{Sorella:2001}. The large improvement of the above wave 
function with respect to  the simpler
one used in Refs.~\cite{Gandolfi:2007,Gandolfi:2007b} is confirmed by
the fact that the variational energies for  both nuclei and symmetric
nuclear matter are negative. This is not true for the
simple wave functions  without tensor correlations.

The mean-field wave function $\Phi(R,S)=\langle RS|\Phi\rangle$ has
the proper quantum numbers and asymptotic behavior.  Its general
form is given by a sum of Slater determinants of the form ${\cal
A}\{\phi_\alpha(\boldsymbol{r_i},s_i)\} \,,$ where $\cal{A}$ is the
antisymmetrizer operator, $\phi_\alpha$ are single particle orbitals
that have the proper asymptotic behavior depending on the system,
and $\alpha$ are the single-particle quantum numbers. For the case
of nuclei, a sum of many Slater determinants is sometimes needed to
give the correct quantum numbers $(\pi,J,T)$ for the nucleus of interest.
The spatial orbital forms are obtained from a Hartree-Fock calculation
with Skyrme forces. The form is described in Ref.~\cite{Maris:2013b}
with the addition of the isospin. In the case of nuclear matter, the
spatial parts of $\phi_\alpha$ are plane-waves with momenta fitting the
simulation box as described in Ref.~\cite{Gandolfi:2009}. Note that the
inclusion of BCS correlations is straightforward; it is easy to replace
$\Phi(R,S)$ with a BCS form written as a Pfaffian as in superfluid neutron
matter~\cite{Gandolfi:2008,Gandolfi:2009b}. Pairing correlations are
expected to be important in describing even-odd splittings in open-shell
nuclei, in neutron-rich nuclei, and in nuclear matter at lower densities.

\begin{table}[h]

\begin{tabular}{lcc}
\hline
Hamiltonian            & AFDMC & GFMC \\
\hline
\hline
AV6$'$                 & -27.09(3) & -26.85(2)   \\
AV7$'$                 & -25.7(2)  &  -26.2(1) \\
N$^2$LO (R$_0$=1.0 fm) & -24.41(3) & -24.56(1) \\
N$^2$LO (R$_0$=1.2 fm) & -25.77(2) & -25.75(1) \\
\hline
\hline
\end{tabular}
\caption{Binding energies for $^4$He using different
two-body interactions. The GFMC energies are taken from 
Refs.~\cite{Lynn:2014,Wiringa:2002}.
The Coulomb contribution has been 
perturbatively subtracted from GFMC results.
}
\label{tab:alpha}
\end{table}

In order to demonstrate that our results are accurate for asymmetric
matter, we first show some results for light nuclei where accurate GFMC
calculations are available. We then show results
for some medium-size nuclei with comparison to experiment.  Having
demonstrated the accuracy of the AFDMC method with the wave function of
Eq.~\ref{eq.wf}, we show results for asymmetric nuclear matter.

We have calculated the binding energies of $^4$He using AV6$'$, AV7$'$,
and the chiral N$^2$LO, and compared those with the corresponding
GFMC values taken from Ref.~\cite{Lynn:2014,Wiringa:2002}. As shown
in table~\ref{tab:alpha}, all the results are in good agreement; the
difference between AFDMC and GFMC is less than 0.125 MeV per nucleon.
We have also compared GFMC and AFDMC results for $^4$He using chiral
forces at NLO and LO, getting a similar agreement; these study will be
the subject of a following paper.

We have also calculated the energy of $^6$Li using the AV6$'$ potential. The
physical structure of this nucleus is complicated, and the results
using GFMC have been obtained by including all the possible spacial/spin
symmetries in s- and p-wave orbitals in the variational wave function, as
well as cluster-dependent two-nucleon correlations~\cite{Pudliner:1997}.
We have implemented a much simpler variational wave function of the
form of Eq. \ref{eq.wf} using a $jj$ basis. The energy obtained
with AFDMC is -28.9(2) MeV compared to the -29.57(4) of
GFMC (subtracting the EM contributions). Since $^6$Li is one of the
most challenging systems to test the accuracy of AFDMC, the results
obtained with this simple wave function are very encouraging. Other
light nuclei have important clustering effects, and they will require
more sophisticated variational wave functions to be implemented in AFDMC
calculations, beyond the scope of this paper.

\begin{table}[h]
\begin{tabular}{lccc}
\hline
                       & AV6$'$ & AV7$'$ & exp\\
\hline
\hline
$^4$He                 & -27.09(3) & -25.7(2)  &  -28.295\\
$^{16}$O               & -115.6(3) & -90.6(4)  &  -127.619\\
$^{40}$Ca              & -322(2)   & -209(1)   &  -342.051\\
\hline
\hline
\end{tabular}
\caption{Binding energy of $^{16}$O and $^{40}$Ca using Argonne NN forces. 
The experimental energies are also shown.}
\label{tab:nuclei}
\end{table}

Using the same Argonne NN forces, we have calculated the
ground-state energy of $^{16}$O and $^{40}$Ca. The results are
shown in table~\ref{tab:nuclei}.  By comparing the results with the
experimental data, it is clear that both NN Hamiltonians underbind
these nuclei, as is the case of $^{4}$He.  We may conclude that
using Argonne  AV6$'$ and AV7$'$ NN forces, the (missing) three-body
force should be attractive. This will need further investigation,
but already shows interesting features.  Using chiral forces
in coupled cluster calculations, the three-body is attractive
in the case of $^{16}$O~\cite{Hagen:2012b}, and repulsive in
$^{40}$Ca~\cite{Hagen:2012}. Within in-medium SRG approach,
the three-body force is attractive for several nuclei from A=4
to 56~\cite{Hergert:2013}. Finally,  SCGF calculations of oxygen,
nitrogen and fluorine isotopes indicate that the three-body force is
attractive~\cite{Cipollone:2013}.  Other recent coupled-cluster
results have been obtained by also including few-body nuclei
when fitting the NN potential. In this case, the contribution
required from three-body forces for medium mass nuclei seems to be very
small~\cite{Ekstrom:2013}.

\begin{figure}[th]
\includegraphics[width=0.9\columnwidth]{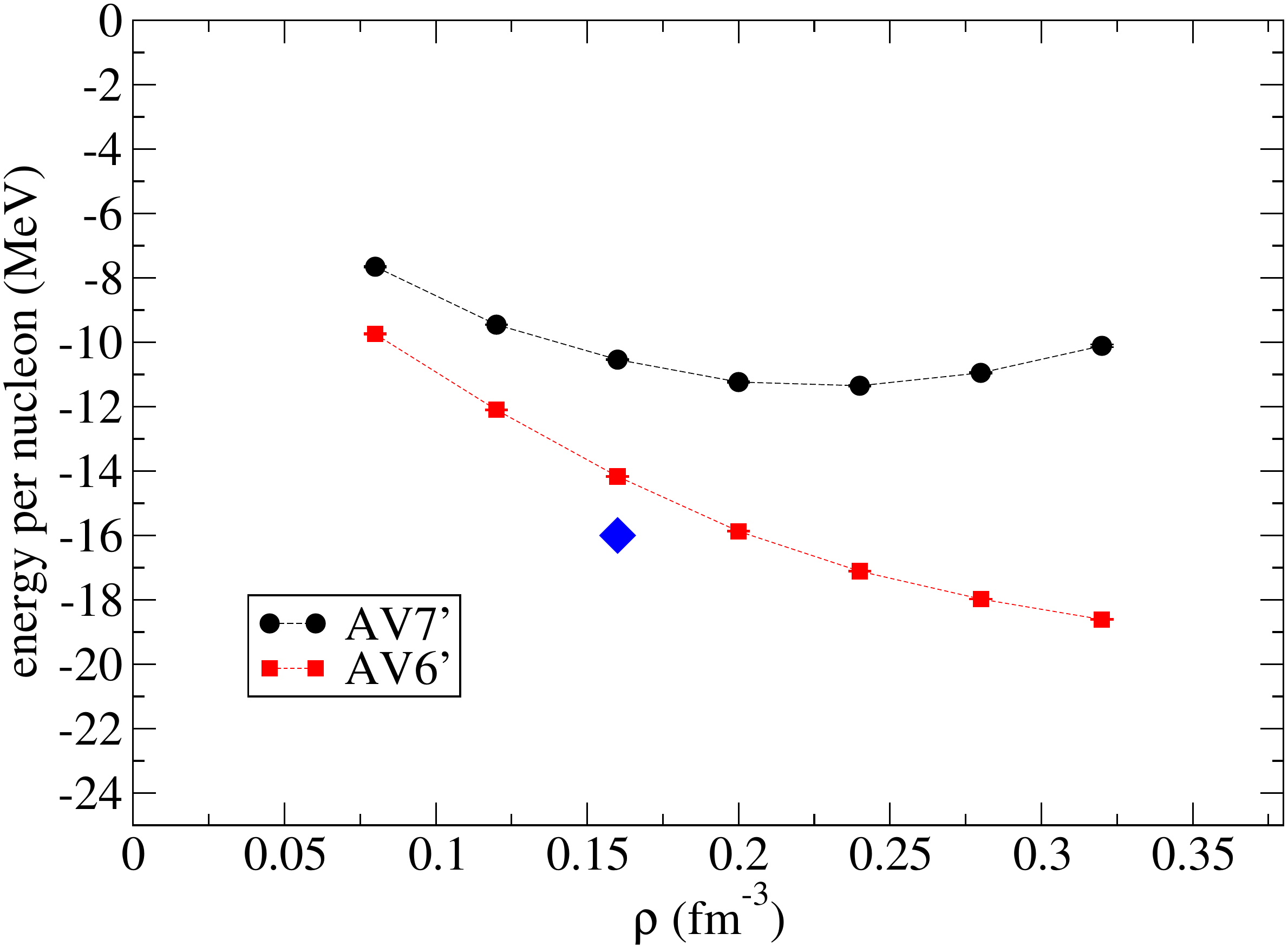}
\caption{(color online) The energy per nucleon of symmetric nuclear matter
obtained using the Argonne AV6$'$ and AV7$'$ interactions,
see the text for details. The blue
diamond indicates the hypothetical saturation point of nuclear matter.}
\label{fig:eos}
\end{figure}

We have also calculated the EOS of symmetric nuclear matter using both
the Argonne AV6$'$ and AV7$'$ interactions, with results shown in
Fig.~\ref{fig:eos}. We simulated infinite matter using 28 nucleons
in a periodic box. Finite size corrections due to the truncation
of the pairwise potential to distances equal to $L/2$  ($L$ is the
box size) have been included as described in Ref.~\cite{Sarsa:2003}.
We have also calculated the energy given by 76, 108, and 132 nucleons at
$\rho$=0.16 fm$^{-3}$, the results are -14.16(2), -13.91(2) and -12.98(4)
respectively, compared to -14.17(2) for 28 nucleons. As expected, the
energy for the larger systems is higher, consistent with
the fact that the trial wave function used for the path constraint is
not extensive.

We have tested the accuracy of AFDMC for nuclear matter by changing
the variational parameters of the spin/isospin dependent correlations. We found
that even in the case where the VMC energy is not optimal, the
AFDMC results  are consistent within statistical errors. We have
included backflow correlations in $\Phi(R,S)$, as commonly done in
liquid atomic $^3$He\cite{Schmidt:1981,Lee:1981} and the electron
gas\cite{Holtzmann:2003}, where it is motivated by the repulsive part of
the two-body potential. 
We have checked that backflow correlations produce the same AFDMC energies
within statistical error bars.

As is clear from Fig.~\ref{fig:eos}, the two different Argonne NN
potentials give quite different results, in particular different
saturation densities. This is due to the fact that the AV6$'$
and AV7$'$ interactions have different nucleon-nucleon phase
shifts~\cite{supplemental}. In any case, comparing with the energy
at $\rho$=0.16 fm$^{-3}$ extracted from heavy nuclei, (in the figure
the blue diamond correspond to the saturation point), it is clear that
both NN Hamiltonians underbind nuclear matter. This is consistent with
the results of $^{16}$O and $^{40}$Ca shown in table~\ref{tab:nuclei}.
The spin/isospin structure of Argonne AV7$'$ is the same as 
local chiral forces of Ref.~\cite{Gezerlis:2013}; their implementation
in the AFDMC method is straightforward. Some preliminary calculations
show that, using N$^2$LO with different cutoffs, the spread of the energy of nuclear
matter is similar to the difference between
AV6$'$ and AV7$'$. A
detailed analysis of the EOS calculated using chiral forces will be the
performed in a future work.

\begin{figure}[th]
\includegraphics[width=0.9\columnwidth]{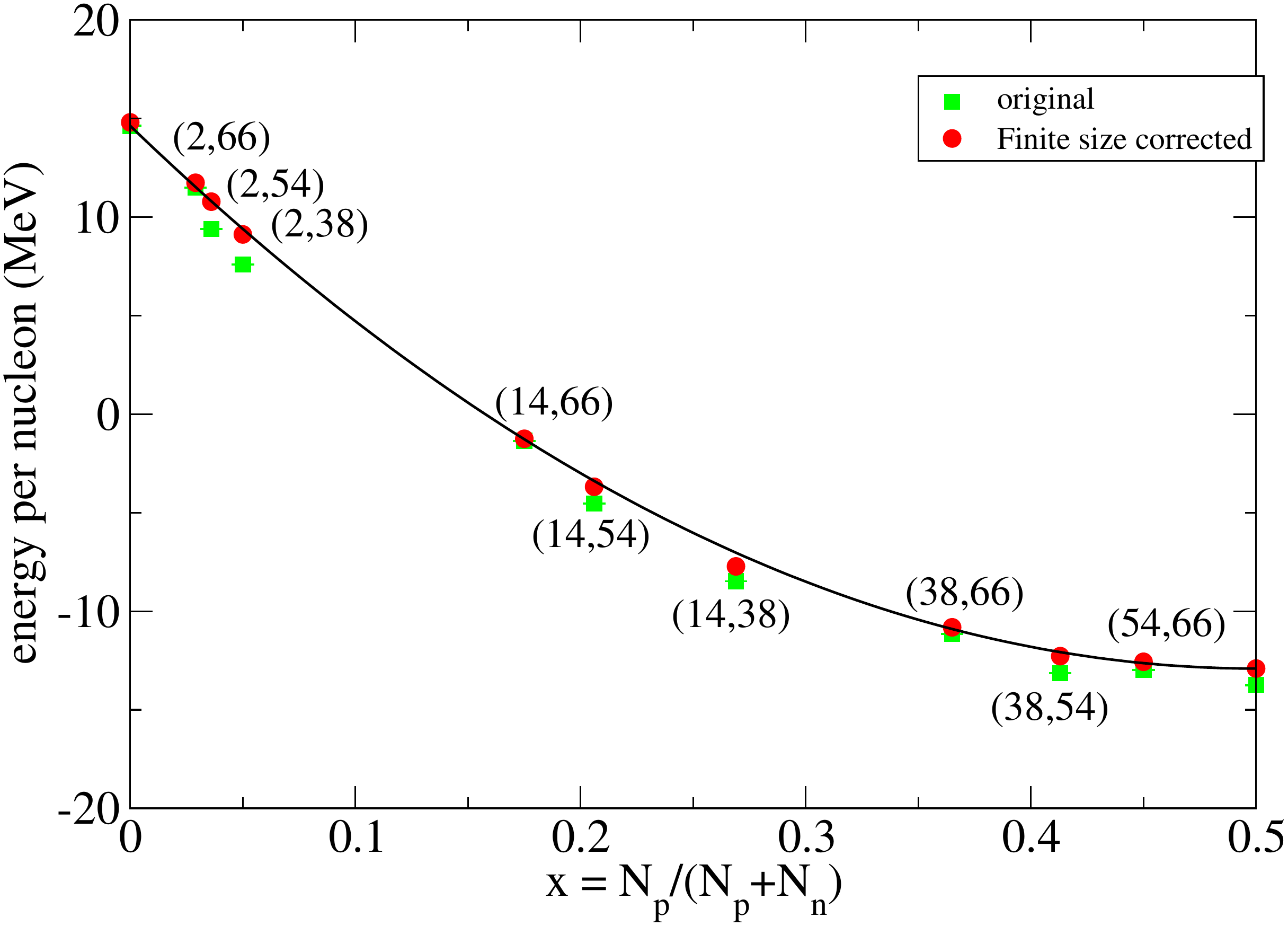}
\caption{(color online) Energy per nucleon of isospin asymmetric nuclear
matter calculated using AV6$'$ potential as a function of the proton
concentration.  Green squares represent the AFDMC results and red circles
the ones in which finite size corrections are included. The numbers
in parenthesis indicate the number of  protons and neutrons considered
in the simulations.}
\label{fig:esym}
\end{figure}

We have calculated the energy of isospin-asymmetric nuclear matter at
$\rho$=0.16 fm$^{-3}$ using the AV6$'$ interaction. We performed simulations
using different combinations of neutrons and protons, listed in 
Fig.~\ref{fig:esym}, filling closed shells of the discretised momentum. 
Corrections for the finite size effects due to the interaction are included as
described in~\cite{Sarsa:2003}. In order to alleviate finite size effects arising 
from the kinetic energy, we have corrected the AFDMC energies by subtracting 
from the AFDMC results the term
\begin{equation}
\delta E(\rho)=E_0(N_n,N_p,\rho)-E_{FG}(p,\rho) \,.
\end{equation}
In the previous equation, $E_0$ is the energy of non-interacting $N_n$
neutrons and $N_p$ protons in the same simulation box and $E_{FG}$
is the energy in the thermodynamic limit at the same isospin polarization,
i.e. $E_{FG}(p,\rho)=E_{FG}(\rho)((1+p)^{5/3}+(1-p)^{5/3})/2$.  This
strategy has been successfully applied to study strongly interacting
polarized Fermi liquids~\cite{Tanatar:1989,Pilati:private,Pilati:2010}.

From Fig.~\ref{fig:esym} we see that our results agree with the quadratic
behavior of the energy as a function of the isospin-asymmetry obtained
by simply interpolating the results for $x=0$ (pure neutron matter) and
$x=0.5$ (symmetric nuclear matter). We do not
expect that using the AV7$'$ interaction the quadratic behavior of the
energy as a function of the asymmetry would change. However, pairing
correlations might play an important  role, especially at lower densities,
and will be included and addressed in future works. It will be interesting
to see if the inclusion of superfluid pairing alters these results.

Clearly it is important to include three-nucleon interactions in
AFDMC. For pure neutron systems, three-body forces can be included
exactly in the propagator because the spin/isospin operators reduce to
a quadratic form in the spin~\cite{Sarsa:2003}. In the case of nuclei
and nuclear matter, the full three-body force cannot yet be included in
the propagator. However, it is possible to use
a simplified form of the three-body force compatible with standard
AFDMC and calculate the difference from the full three-body potential
as a perturbation.  This strategy has been extensively tested in GFMC
calculations~\cite{Pudliner:1997}. Another approach consists in reducing
the three-body potential to a $V_2(\rho)$ density dependent force as
done in Ref.~\cite{Lovato:2011}, and perturbatively compute the difference
$[V_2(\rho)-V_3]$.

In conclusion, we have presented a new AFDMC method extended to NN
forces that include spin-orbit terms, along with a significantly improved
variational wave function and improved propagation technique. 
Since the forces have the same spin/isospin
operatorial structure of local chiral forces at N$^2$LO, the extension
of the AFDMC to use them is straightforward, similar to what has been
done for pure neutron matter~\cite{Gezerlis:2013,Gezerlis:2014}.
We have also presented the first quantum Monte Carlo calculation of
asymmetric nuclear matter using bare NN nuclear interactions, showing that
at saturation density the energy per particle follows the often assumed
quadratic behavior as a function of isospin asymmetry.  This work paves
the way for a systematic study of the structure of  medium mass nuclei,
neutron-rich nuclei, and nuclear matter using both Argonne and chiral
forces with unprecedented accuracy.

\acknowledgments{
We would like to thank P. Armani for very important comments and
suggestions in the very early stage of this work, and D.
Lonardoni, J. Lynn, S.C. Pieper, S. Pilati, and R.B. Wiringa for the many
useful discussions. 
Computer time was provided by Los Alamos Institutional Computing and
by an INCITE allocation at ANL. This research used resources of the
National Energy Research Scientific Computing Center, which is supported
by the Office of Science of the U.S. Department of Energy under Contract
No. DE-AC02-05CH11231. The work of J. Carlson and S. Gandolfi was
supported by the Department of Energy Nuclear Physics Office, and by the
NUCLEI SciDAC program. A. Lovato was supported by the Department of Energy, 
Office of Nuclear Physics, under contract No. DE-AC02-06CH11357
K. Schmidt was supported by the National Science Foundation grant PHY-1067777. 
}


\begin{thebibliography}{56}%
\makeatletter
\providecommand \@ifxundefined [1]{%
 \@ifx{#1\undefined}
}%
\providecommand \@ifnum [1]{%
 \ifnum #1\expandafter \@firstoftwo
 \else \expandafter \@secondoftwo
 \fi
}%
\providecommand \@ifx [1]{%
 \ifx #1\expandafter \@firstoftwo
 \else \expandafter \@secondoftwo
 \fi
}%
\providecommand \natexlab [1]{#1}%
\providecommand \enquote  [1]{``#1''}%
\providecommand \bibnamefont  [1]{#1}%
\providecommand \bibfnamefont [1]{#1}%
\providecommand \citenamefont [1]{#1}%
\providecommand \href@noop [0]{\@secondoftwo}%
\providecommand \href [0]{\begingroup \@sanitize@url \@href}%
\providecommand \@href[1]{\@@startlink{#1}\@@href}%
\providecommand \@@href[1]{\endgroup#1\@@endlink}%
\providecommand \@sanitize@url [0]{\catcode `\\12\catcode `\$12\catcode
  `\&12\catcode `\#12\catcode `\^12\catcode `\_12\catcode `\%12\relax}%
\providecommand \@@startlink[1]{}%
\providecommand \@@endlink[0]{}%
\providecommand \url  [0]{\begingroup\@sanitize@url \@url }%
\providecommand \@url [1]{\endgroup\@href {#1}{\urlprefix }}%
\providecommand \urlprefix  [0]{URL }%
\providecommand \Eprint [0]{\href }%
\providecommand \doibase [0]{http://dx.doi.org/}%
\providecommand \selectlanguage [0]{\@gobble}%
\providecommand \bibinfo  [0]{\@secondoftwo}%
\providecommand \bibfield  [0]{\@secondoftwo}%
\providecommand \translation [1]{[#1]}%
\providecommand \BibitemOpen [0]{}%
\providecommand \bibitemStop [0]{}%
\providecommand \bibitemNoStop [0]{.\EOS\space}%
\providecommand \EOS [0]{\spacefactor3000\relax}%
\providecommand \BibitemShut  [1]{\csname bibitem#1\endcsname}%
\let\auto@bib@innerbib\@empty
\bibitem [{\citenamefont {Wiringa}\ \emph {et~al.}(1995)\citenamefont
  {Wiringa}, \citenamefont {Stoks},\ and\ \citenamefont
  {Schiavilla}}]{Wiringa:1995}%
  \BibitemOpen
  \bibfield  {author} {\bibinfo {author} {\bibfnamefont {R.~B.}\ \bibnamefont
  {Wiringa}}, \bibinfo {author} {\bibfnamefont {V.~G.~J.}\ \bibnamefont
  {Stoks}}, \ and\ \bibinfo {author} {\bibfnamefont {R.}~\bibnamefont
  {Schiavilla}},\ }\href {\doibase 10.1103/PhysRevC.51.38} {\bibfield
  {journal} {\bibinfo  {journal} {Phys. Rev. C}\ }\textbf {\bibinfo {volume}
  {51}},\ \bibinfo {pages} {38} (\bibinfo {year} {1995})}\BibitemShut {NoStop}%
\bibitem [{\citenamefont {Machleidt}(2001)}]{Machleidt:2001}%
  \BibitemOpen
  \bibfield  {author} {\bibinfo {author} {\bibfnamefont {R.}~\bibnamefont
  {Machleidt}},\ }\href {\doibase 10.1103/PhysRevC.63.024001} {\bibfield
  {journal} {\bibinfo  {journal} {Phys. Rev. C}\ }\textbf {\bibinfo {volume}
  {63}},\ \bibinfo {pages} {024001} (\bibinfo {year} {2001})}\BibitemShut
  {NoStop}%
\bibitem [{\citenamefont {Entem}\ and\ \citenamefont
  {Machleidt}(2003)}]{Entem:2003}%
  \BibitemOpen
  \bibfield  {author} {\bibinfo {author} {\bibfnamefont {D.~R.}\ \bibnamefont
  {Entem}}\ and\ \bibinfo {author} {\bibfnamefont {R.}~\bibnamefont
  {Machleidt}},\ }\href {\doibase 10.1103/PhysRevC.68.041001} {\bibfield
  {journal} {\bibinfo  {journal} {Phys. Rev. C}\ }\textbf {\bibinfo {volume}
  {68}},\ \bibinfo {pages} {041001} (\bibinfo {year} {2003})}\BibitemShut
  {NoStop}%
\bibitem [{\citenamefont {Pieper}\ \emph {et~al.}(2001)\citenamefont {Pieper},
  \citenamefont {Pandharipande}, \citenamefont {Wiringa},\ and\ \citenamefont
  {Carlson}}]{Pieper:2001}%
  \BibitemOpen
  \bibfield  {author} {\bibinfo {author} {\bibfnamefont {S.~C.}\ \bibnamefont
  {Pieper}}, \bibinfo {author} {\bibfnamefont {V.~R.}\ \bibnamefont
  {Pandharipande}}, \bibinfo {author} {\bibfnamefont {R.~B.}\ \bibnamefont
  {Wiringa}}, \ and\ \bibinfo {author} {\bibfnamefont {J.}~\bibnamefont
  {Carlson}},\ }\href {\doibase 10.1103/PhysRevC.64.014001} {\bibfield
  {journal} {\bibinfo  {journal} {Phys. Rev. C}\ }\textbf {\bibinfo {volume}
  {64}},\ \bibinfo {pages} {014001} (\bibinfo {year} {2001})}\BibitemShut
  {NoStop}%
\bibitem [{\citenamefont {Maris}\ \emph
  {et~al.}(2013{\natexlab{a}})\citenamefont {Maris}, \citenamefont {Vary},\
  and\ \citenamefont {Navr\'atil}}]{Maris:2013}%
  \BibitemOpen
  \bibfield  {author} {\bibinfo {author} {\bibfnamefont {P.}~\bibnamefont
  {Maris}}, \bibinfo {author} {\bibfnamefont {J.~P.}\ \bibnamefont {Vary}}, \
  and\ \bibinfo {author} {\bibfnamefont {P.}~\bibnamefont {Navr\'atil}},\
  }\href {\doibase 10.1103/PhysRevC.87.014327} {\bibfield  {journal} {\bibinfo
  {journal} {Phys. Rev. C}\ }\textbf {\bibinfo {volume} {87}},\ \bibinfo
  {pages} {014327} (\bibinfo {year} {2013}{\natexlab{a}})}\BibitemShut
  {NoStop}%
\bibitem [{\citenamefont {Hagen}\ \emph
  {et~al.}(2012{\natexlab{a}})\citenamefont {Hagen}, \citenamefont
  {Hjorth-Jensen}, \citenamefont {Jansen}, \citenamefont {Machleidt},\ and\
  \citenamefont {Papenbrock}}]{Hagen:2013}%
  \BibitemOpen
  \bibfield  {author} {\bibinfo {author} {\bibfnamefont {G.}~\bibnamefont
  {Hagen}}, \bibinfo {author} {\bibfnamefont {M.}~\bibnamefont
  {Hjorth-Jensen}}, \bibinfo {author} {\bibfnamefont {G.~R.}\ \bibnamefont
  {Jansen}}, \bibinfo {author} {\bibfnamefont {R.}~\bibnamefont {Machleidt}}, \
  and\ \bibinfo {author} {\bibfnamefont {T.}~\bibnamefont {Papenbrock}},\
  }\href {\doibase 10.1103/PhysRevLett.109.032502} {\bibfield  {journal}
  {\bibinfo  {journal} {Phys. Rev. Lett.}\ }\textbf {\bibinfo {volume} {109}},\
  \bibinfo {pages} {032502} (\bibinfo {year} {2012}{\natexlab{a}})}\BibitemShut
  {NoStop}%
\bibitem [{\citenamefont {Binder}\ \emph {et~al.}(2013)\citenamefont {Binder},
  \citenamefont {Langhammer}, \citenamefont {Calci}, \citenamefont
  {Navr\'atil},\ and\ \citenamefont {Roth}}]{Binder:2013}%
  \BibitemOpen
  \bibfield  {author} {\bibinfo {author} {\bibfnamefont {S.}~\bibnamefont
  {Binder}}, \bibinfo {author} {\bibfnamefont {J.}~\bibnamefont {Langhammer}},
  \bibinfo {author} {\bibfnamefont {A.}~\bibnamefont {Calci}}, \bibinfo
  {author} {\bibfnamefont {P.}~\bibnamefont {Navr\'atil}}, \ and\ \bibinfo
  {author} {\bibfnamefont {R.}~\bibnamefont {Roth}},\ }\href {\doibase
  10.1103/PhysRevC.87.021303} {\bibfield  {journal} {\bibinfo  {journal} {Phys.
  Rev. C}\ }\textbf {\bibinfo {volume} {87}},\ \bibinfo {pages} {021303}
  (\bibinfo {year} {2013})}\BibitemShut {NoStop}%
\bibitem [{\citenamefont {Gandolfi}\ \emph
  {et~al.}(2009{\natexlab{a}})\citenamefont {Gandolfi}, \citenamefont
  {Illarionov}, \citenamefont {Schmidt}, \citenamefont {Pederiva},\ and\
  \citenamefont {Fantoni}}]{Gandolfi:2009}%
  \BibitemOpen
  \bibfield  {author} {\bibinfo {author} {\bibfnamefont {S.}~\bibnamefont
  {Gandolfi}}, \bibinfo {author} {\bibfnamefont {A.~Y.}\ \bibnamefont
  {Illarionov}}, \bibinfo {author} {\bibfnamefont {K.~E.}\ \bibnamefont
  {Schmidt}}, \bibinfo {author} {\bibfnamefont {F.}~\bibnamefont {Pederiva}}, \
  and\ \bibinfo {author} {\bibfnamefont {S.}~\bibnamefont {Fantoni}},\ }\href
  {\doibase 10.1103/PhysRevC.79.054005} {\bibfield  {journal} {\bibinfo
  {journal} {Phys. Rev. C}\ }\textbf {\bibinfo {volume} {79}},\ \bibinfo
  {pages} {054005} (\bibinfo {year} {2009}{\natexlab{a}})}\BibitemShut
  {NoStop}%
\bibitem [{\citenamefont {{Gandolfi}}\ \emph {et~al.}(2012)\citenamefont
  {{Gandolfi}}, \citenamefont {{Carlson}},\ and\ \citenamefont
  {{Reddy}}}]{Gandolfi:2012}%
  \BibitemOpen
  \bibfield  {author} {\bibinfo {author} {\bibfnamefont {S.}~\bibnamefont
  {{Gandolfi}}}, \bibinfo {author} {\bibfnamefont {J.}~\bibnamefont
  {{Carlson}}}, \ and\ \bibinfo {author} {\bibfnamefont {S.}~\bibnamefont
  {{Reddy}}},\ }\href {\doibase 10.1103/PhysRevC.85.032801} {\bibfield
  {journal} {\bibinfo  {journal} {\prc}\ }\textbf {\bibinfo {volume} {85}},\
  \bibinfo {eid} {032801} (\bibinfo {year} {2012})}\BibitemShut {NoStop}%
\bibitem [{\citenamefont {Hebeler}\ and\ \citenamefont
  {Schwenk}(2010)}]{Hebeler:2010}%
  \BibitemOpen
  \bibfield  {author} {\bibinfo {author} {\bibfnamefont {K.}~\bibnamefont
  {Hebeler}}\ and\ \bibinfo {author} {\bibfnamefont {A.}~\bibnamefont
  {Schwenk}},\ }\href {\doibase 10.1103/PhysRevC.82.014314} {\bibfield
  {journal} {\bibinfo  {journal} {Phys. Rev. C}\ }\textbf {\bibinfo {volume}
  {82}},\ \bibinfo {pages} {014314} (\bibinfo {year} {2010})}\BibitemShut
  {NoStop}%
\bibitem [{\citenamefont {Carlson}(1987)}]{Carlson:1987}%
  \BibitemOpen
  \bibfield  {author} {\bibinfo {author} {\bibfnamefont {J.}~\bibnamefont
  {Carlson}},\ }\href {\doibase 10.1103/PhysRevC.36.2026} {\bibfield  {journal}
  {\bibinfo  {journal} {Phys. Rev. C}\ }\textbf {\bibinfo {volume} {36}},\
  \bibinfo {pages} {2026} (\bibinfo {year} {1987})}\BibitemShut {NoStop}%
\bibitem [{\citenamefont {Navr\'atil}\ \emph {et~al.}(2000)\citenamefont
  {Navr\'atil}, \citenamefont {Vary},\ and\ \citenamefont
  {Barrett}}]{Navratil:2000}%
  \BibitemOpen
  \bibfield  {author} {\bibinfo {author} {\bibfnamefont {P.}~\bibnamefont
  {Navr\'atil}}, \bibinfo {author} {\bibfnamefont {J.~P.}\ \bibnamefont
  {Vary}}, \ and\ \bibinfo {author} {\bibfnamefont {B.~R.}\ \bibnamefont
  {Barrett}},\ }\href {\doibase 10.1103/PhysRevLett.84.5728} {\bibfield
  {journal} {\bibinfo  {journal} {Phys. Rev. Lett.}\ }\textbf {\bibinfo
  {volume} {84}},\ \bibinfo {pages} {5728} (\bibinfo {year}
  {2000})}\BibitemShut {NoStop}%
\bibitem [{\citenamefont {Maris}\ \emph {et~al.}(2009)\citenamefont {Maris},
  \citenamefont {Vary},\ and\ \citenamefont {Shirokov}}]{Maris:2009}%
  \BibitemOpen
  \bibfield  {author} {\bibinfo {author} {\bibfnamefont {P.}~\bibnamefont
  {Maris}}, \bibinfo {author} {\bibfnamefont {J.~P.}\ \bibnamefont {Vary}}, \
  and\ \bibinfo {author} {\bibfnamefont {A.~M.}\ \bibnamefont {Shirokov}},\
  }\href {\doibase 10.1103/PhysRevC.79.014308} {\bibfield  {journal} {\bibinfo
  {journal} {Phys. Rev. C}\ }\textbf {\bibinfo {volume} {79}},\ \bibinfo
  {pages} {014308} (\bibinfo {year} {2009})}\BibitemShut {NoStop}%
\bibitem [{\citenamefont {Barnea}\ \emph {et~al.}(1999)\citenamefont {Barnea},
  \citenamefont {Leidemann},\ and\ \citenamefont {Orlandini}}]{Barnea:1999}%
  \BibitemOpen
  \bibfield  {author} {\bibinfo {author} {\bibfnamefont {N.}~\bibnamefont
  {Barnea}}, \bibinfo {author} {\bibfnamefont {W.}~\bibnamefont {Leidemann}}, \
  and\ \bibinfo {author} {\bibfnamefont {G.}~\bibnamefont {Orlandini}},\ }\href
  {\doibase http://dx.doi.org/10.1016/S0375-9474(99)00113-X} {\bibfield
  {journal} {\bibinfo  {journal} {Nuclear Physics A}\ }\textbf {\bibinfo
  {volume} {650}},\ \bibinfo {pages} {427 } (\bibinfo {year}
  {1999})}\BibitemShut {NoStop}%
\bibitem [{\citenamefont {Dean}\ and\ \citenamefont
  {Hjorth-Jensen}(2004)}]{Dean:2004}%
  \BibitemOpen
  \bibfield  {author} {\bibinfo {author} {\bibfnamefont {D.~J.}\ \bibnamefont
  {Dean}}\ and\ \bibinfo {author} {\bibfnamefont {M.}~\bibnamefont
  {Hjorth-Jensen}},\ }\href {\doibase 10.1103/PhysRevC.69.054320} {\bibfield
  {journal} {\bibinfo  {journal} {Phys. Rev. C}\ }\textbf {\bibinfo {volume}
  {69}},\ \bibinfo {pages} {054320} (\bibinfo {year} {2004})}\BibitemShut
  {NoStop}%
\bibitem [{\citenamefont {Dickhoff}\ and\ \citenamefont
  {Barbieri}(2004)}]{Dickhoff:2004}%
  \BibitemOpen
  \bibfield  {author} {\bibinfo {author} {\bibfnamefont {W.}~\bibnamefont
  {Dickhoff}}\ and\ \bibinfo {author} {\bibfnamefont {C.}~\bibnamefont
  {Barbieri}},\ }\href {\doibase http://dx.doi.org/10.1016/j.ppnp.2004.02.038}
  {\bibfield  {journal} {\bibinfo  {journal} {Progress in Particle and Nuclear
  Physics}\ }\textbf {\bibinfo {volume} {52}},\ \bibinfo {pages} {377 }
  (\bibinfo {year} {2004})}\BibitemShut {NoStop}%
\bibitem [{\citenamefont {Tsukiyama}\ \emph {et~al.}(2011)\citenamefont
  {Tsukiyama}, \citenamefont {Bogner},\ and\ \citenamefont
  {Schwenk}}]{Tsukiyama:2011}%
  \BibitemOpen
  \bibfield  {author} {\bibinfo {author} {\bibfnamefont {K.}~\bibnamefont
  {Tsukiyama}}, \bibinfo {author} {\bibfnamefont {S.~K.}\ \bibnamefont
  {Bogner}}, \ and\ \bibinfo {author} {\bibfnamefont {A.}~\bibnamefont
  {Schwenk}},\ }\href {\doibase 10.1103/PhysRevLett.106.222502} {\bibfield
  {journal} {\bibinfo  {journal} {Phys. Rev. Lett.}\ }\textbf {\bibinfo
  {volume} {106}},\ \bibinfo {pages} {222502} (\bibinfo {year}
  {2011})}\BibitemShut {NoStop}%
\bibitem [{\citenamefont {Hebeler}\ \emph {et~al.}(2011)\citenamefont
  {Hebeler}, \citenamefont {Bogner}, \citenamefont {Furnstahl}, \citenamefont
  {Nogga},\ and\ \citenamefont {Schwenk}}]{Hebeler:2011}%
  \BibitemOpen
  \bibfield  {author} {\bibinfo {author} {\bibfnamefont {K.}~\bibnamefont
  {Hebeler}}, \bibinfo {author} {\bibfnamefont {S.~K.}\ \bibnamefont {Bogner}},
  \bibinfo {author} {\bibfnamefont {R.~J.}\ \bibnamefont {Furnstahl}}, \bibinfo
  {author} {\bibfnamefont {A.}~\bibnamefont {Nogga}}, \ and\ \bibinfo {author}
  {\bibfnamefont {A.}~\bibnamefont {Schwenk}},\ }\href {\doibase
  10.1103/PhysRevC.83.031301} {\bibfield  {journal} {\bibinfo  {journal} {Phys.
  Rev. C}\ }\textbf {\bibinfo {volume} {83}},\ \bibinfo {pages} {031301}
  (\bibinfo {year} {2011})}\BibitemShut {NoStop}%
\bibitem [{\citenamefont {Baardsen}\ \emph {et~al.}(2013)\citenamefont
  {Baardsen}, \citenamefont {Ekstr\"om}, \citenamefont {Hagen},\ and\
  \citenamefont {Hjorth-Jensen}}]{Baardsen:2013}%
  \BibitemOpen
  \bibfield  {author} {\bibinfo {author} {\bibfnamefont {G.}~\bibnamefont
  {Baardsen}}, \bibinfo {author} {\bibfnamefont {A.}~\bibnamefont {Ekstr\"om}},
  \bibinfo {author} {\bibfnamefont {G.}~\bibnamefont {Hagen}}, \ and\ \bibinfo
  {author} {\bibfnamefont {M.}~\bibnamefont {Hjorth-Jensen}},\ }\href {\doibase
  10.1103/PhysRevC.88.054312} {\bibfield  {journal} {\bibinfo  {journal} {Phys.
  Rev. C}\ }\textbf {\bibinfo {volume} {88}},\ \bibinfo {pages} {054312}
  (\bibinfo {year} {2013})}\BibitemShut {NoStop}%
\bibitem [{\citenamefont {Hagen}\ \emph {et~al.}(2014)\citenamefont {Hagen},
  \citenamefont {Papenbrock}, \citenamefont {Ekstr\"om}, \citenamefont {Wendt},
  \citenamefont {Baardsen}, \citenamefont {Gandolfi}, \citenamefont
  {Hjorth-Jensen},\ and\ \citenamefont {Horowitz}}]{Hagen:2014}%
  \BibitemOpen
  \bibfield  {author} {\bibinfo {author} {\bibfnamefont {G.}~\bibnamefont
  {Hagen}}, \bibinfo {author} {\bibfnamefont {T.}~\bibnamefont {Papenbrock}},
  \bibinfo {author} {\bibfnamefont {A.}~\bibnamefont {Ekstr\"om}}, \bibinfo
  {author} {\bibfnamefont {K.~A.}\ \bibnamefont {Wendt}}, \bibinfo {author}
  {\bibfnamefont {G.}~\bibnamefont {Baardsen}}, \bibinfo {author}
  {\bibfnamefont {S.}~\bibnamefont {Gandolfi}}, \bibinfo {author}
  {\bibfnamefont {M.}~\bibnamefont {Hjorth-Jensen}}, \ and\ \bibinfo {author}
  {\bibfnamefont {C.~J.}\ \bibnamefont {Horowitz}},\ }\href {\doibase
  10.1103/PhysRevC.89.014319} {\bibfield  {journal} {\bibinfo  {journal} {Phys.
  Rev. C}\ }\textbf {\bibinfo {volume} {89}},\ \bibinfo {pages} {014319}
  (\bibinfo {year} {2014})}\BibitemShut {NoStop}%
\bibitem [{\citenamefont {Schmidt}\ and\ \citenamefont
  {Fantoni}(1999)}]{Schmidt:1999}%
  \BibitemOpen
  \bibfield  {author} {\bibinfo {author} {\bibfnamefont {K.~E.}\ \bibnamefont
  {Schmidt}}\ and\ \bibinfo {author} {\bibfnamefont {S.}~\bibnamefont
  {Fantoni}},\ }\href@noop {} {\bibfield  {journal} {\bibinfo  {journal} {Phys.
  Lett. B}\ }\textbf {\bibinfo {volume} {446}},\ \bibinfo {pages} {99}
  (\bibinfo {year} {1999})}\BibitemShut {NoStop}%
\bibitem [{\citenamefont {Pieper}(2005)}]{Pieper:2005}%
  \BibitemOpen
  \bibfield  {author} {\bibinfo {author} {\bibfnamefont {S.~C.}\ \bibnamefont
  {Pieper}},\ }\href {\doibase 10.1016/j.nuclphysa.2005.02.018} {\bibfield
  {journal} {\bibinfo  {journal} {Nuclear Physics A}\ }\textbf {\bibinfo
  {volume} {751}},\ \bibinfo {pages} {516} (\bibinfo {year}
  {2005})}\BibitemShut {NoStop}%
\bibitem [{\citenamefont {Lovato}\ \emph {et~al.}(2013)\citenamefont {Lovato},
  \citenamefont {Gandolfi}, \citenamefont {Butler}, \citenamefont {Carlson},
  \citenamefont {Lusk}, \citenamefont {Pieper},\ and\ \citenamefont
  {Schiavilla}}]{Lovato:2013}%
  \BibitemOpen
  \bibfield  {author} {\bibinfo {author} {\bibfnamefont {A.}~\bibnamefont
  {Lovato}}, \bibinfo {author} {\bibfnamefont {S.}~\bibnamefont {Gandolfi}},
  \bibinfo {author} {\bibfnamefont {R.}~\bibnamefont {Butler}}, \bibinfo
  {author} {\bibfnamefont {J.}~\bibnamefont {Carlson}}, \bibinfo {author}
  {\bibfnamefont {E.}~\bibnamefont {Lusk}}, \bibinfo {author} {\bibfnamefont
  {S.~C.}\ \bibnamefont {Pieper}}, \ and\ \bibinfo {author} {\bibfnamefont
  {R.}~\bibnamefont {Schiavilla}},\ }\href {\doibase
  10.1103/PhysRevLett.111.092501} {\bibfield  {journal} {\bibinfo  {journal}
  {Phys. Rev. Lett.}\ }\textbf {\bibinfo {volume} {111}},\ \bibinfo {pages}
  {092501} (\bibinfo {year} {2013})}\BibitemShut {NoStop}%
\bibitem [{\citenamefont {Lovato}\ \emph {et~al.}(2014)\citenamefont {Lovato},
  \citenamefont {Gandolfi}, \citenamefont {Carlson}, \citenamefont {Pieper},\
  and\ \citenamefont {Schiavilla}}]{Lovato:2014}%
  \BibitemOpen
  \bibfield  {author} {\bibinfo {author} {\bibfnamefont {A.}~\bibnamefont
  {Lovato}}, \bibinfo {author} {\bibfnamefont {S.}~\bibnamefont {Gandolfi}},
  \bibinfo {author} {\bibfnamefont {J.}~\bibnamefont {Carlson}}, \bibinfo
  {author} {\bibfnamefont {S.~C.}\ \bibnamefont {Pieper}}, \ and\ \bibinfo
  {author} {\bibfnamefont {R.}~\bibnamefont {Schiavilla}},\ }\href {\doibase
  10.1103/PhysRevLett.112.182502} {\bibfield  {journal} {\bibinfo  {journal}
  {Phys. Rev. Lett.}\ }\textbf {\bibinfo {volume} {112}},\ \bibinfo {pages}
  {182502} (\bibinfo {year} {2014})}\BibitemShut {NoStop}%
\bibitem [{\citenamefont {Gezerlis}\ \emph {et~al.}(2013)\citenamefont
  {Gezerlis}, \citenamefont {Tews}, \citenamefont {Epelbaum}, \citenamefont
  {Gandolfi}, \citenamefont {Hebeler}, \citenamefont {Nogga},\ and\
  \citenamefont {Schwenk}}]{Gezerlis:2013}%
  \BibitemOpen
  \bibfield  {author} {\bibinfo {author} {\bibfnamefont {A.}~\bibnamefont
  {Gezerlis}}, \bibinfo {author} {\bibfnamefont {I.}~\bibnamefont {Tews}},
  \bibinfo {author} {\bibfnamefont {E.}~\bibnamefont {Epelbaum}}, \bibinfo
  {author} {\bibfnamefont {S.}~\bibnamefont {Gandolfi}}, \bibinfo {author}
  {\bibfnamefont {K.}~\bibnamefont {Hebeler}}, \bibinfo {author} {\bibfnamefont
  {A.}~\bibnamefont {Nogga}}, \ and\ \bibinfo {author} {\bibfnamefont
  {A.}~\bibnamefont {Schwenk}},\ }\href {\doibase
  10.1103/PhysRevLett.111.032501} {\bibfield  {journal} {\bibinfo  {journal}
  {Phys. Rev. Lett.}\ }\textbf {\bibinfo {volume} {111}},\ \bibinfo {pages}
  {032501} (\bibinfo {year} {2013})}\BibitemShut {NoStop}%
\bibitem [{\citenamefont {Lynn}\ \emph {et~al.}()\citenamefont {Lynn},
  \citenamefont {Carlson}, \citenamefont {Epelbaum}, \citenamefont {Gandolfi},
  \citenamefont {Gezerlis},\ and\ \citenamefont {Schwenk}}]{Lynn:2014}%
  \BibitemOpen
  \bibfield  {author} {\bibinfo {author} {\bibfnamefont {J.~E.}\ \bibnamefont
  {Lynn}}, \bibinfo {author} {\bibfnamefont {J.}~\bibnamefont {Carlson}},
  \bibinfo {author} {\bibfnamefont {E.}~\bibnamefont {Epelbaum}}, \bibinfo
  {author} {\bibfnamefont {S.}~\bibnamefont {Gandolfi}}, \bibinfo {author}
  {\bibfnamefont {A.}~\bibnamefont {Gezerlis}}, \ and\ \bibinfo {author}
  {\bibfnamefont {A.}~\bibnamefont {Schwenk}},\ }\href@noop {} {\bibinfo
  {journal} {arXiv:1406.2787}\ }\BibitemShut {NoStop}%
\bibitem [{\citenamefont {Gandolfi}\ \emph
  {et~al.}(2007{\natexlab{a}})\citenamefont {Gandolfi}, \citenamefont
  {Pederiva}, \citenamefont {Fantoni},\ and\ \citenamefont
  {Schmidt}}]{Gandolfi:2007}%
  \BibitemOpen
\bibfield  {journal} {  }\bibfield  {author} {\bibinfo {author} {\bibfnamefont
  {S.}~\bibnamefont {Gandolfi}}, \bibinfo {author} {\bibfnamefont
  {F.}~\bibnamefont {Pederiva}}, \bibinfo {author} {\bibfnamefont
  {S.}~\bibnamefont {Fantoni}}, \ and\ \bibinfo {author} {\bibfnamefont
  {K.~E.}\ \bibnamefont {Schmidt}},\ }\href {\doibase
  10.1103/PhysRevLett.98.102503} {\bibfield  {journal} {\bibinfo  {journal}
  {Phys. Rev. Lett.}\ }\textbf {\bibinfo {volume} {98}},\ \bibinfo {pages}
  {102503} (\bibinfo {year} {2007}{\natexlab{a}})}\BibitemShut {NoStop}%
\bibitem [{\citenamefont {Gandolfi}\ \emph
  {et~al.}(2007{\natexlab{b}})\citenamefont {Gandolfi}, \citenamefont
  {Pederiva}, \citenamefont {Fantoni},\ and\ \citenamefont
  {Schmidt}}]{Gandolfi:2007b}%
  \BibitemOpen
  \bibfield  {author} {\bibinfo {author} {\bibfnamefont {S.}~\bibnamefont
  {Gandolfi}}, \bibinfo {author} {\bibfnamefont {F.}~\bibnamefont {Pederiva}},
  \bibinfo {author} {\bibfnamefont {S.}~\bibnamefont {Fantoni}}, \ and\
  \bibinfo {author} {\bibfnamefont {K.~E.}\ \bibnamefont {Schmidt}},\ }\href
  {\doibase 10.1103/PhysRevLett.99.022507} {\bibfield  {journal} {\bibinfo
  {journal} {Phys. Rev. Lett.}\ }\textbf {\bibinfo {volume} {99}},\ \bibinfo
  {pages} {022507} (\bibinfo {year} {2007}{\natexlab{b}})}\BibitemShut
  {NoStop}%
\bibitem [{\citenamefont {Lagaris}\ and\ \citenamefont
  {Pandharipande}(1981{\natexlab{a}})}]{Lagaris:1981b}%
  \BibitemOpen
  \bibfield  {author} {\bibinfo {author} {\bibfnamefont {I.}~\bibnamefont
  {Lagaris}}\ and\ \bibinfo {author} {\bibfnamefont {V.}~\bibnamefont
  {Pandharipande}},\ }\href {\doibase
  http://dx.doi.org/10.1016/0375-9474(81)90032-4} {\bibfield  {journal}
  {\bibinfo  {journal} {Nuclear Physics A}\ }\textbf {\bibinfo {volume}
  {369}},\ \bibinfo {pages} {470 } (\bibinfo {year}
  {1981}{\natexlab{a}})}\BibitemShut {NoStop}%
\bibitem [{\citenamefont {Zuo}\ \emph {et~al.}(1999)\citenamefont {Zuo},
  \citenamefont {Bombaci},\ and\ \citenamefont {Lombardo}}]{Zuo:1999}%
  \BibitemOpen
  \bibfield  {author} {\bibinfo {author} {\bibfnamefont {W.}~\bibnamefont
  {Zuo}}, \bibinfo {author} {\bibfnamefont {I.}~\bibnamefont {Bombaci}}, \ and\
  \bibinfo {author} {\bibfnamefont {U.}~\bibnamefont {Lombardo}},\ }\href
  {\doibase 10.1103/PhysRevC.60.024605} {\bibfield  {journal} {\bibinfo
  {journal} {Phys. Rev. C}\ }\textbf {\bibinfo {volume} {60}},\ \bibinfo
  {pages} {024605} (\bibinfo {year} {1999})}\BibitemShut {NoStop}%
\bibitem [{\citenamefont {Drischler}\ \emph {et~al.}(2014)\citenamefont
  {Drischler}, \citenamefont {Som\`a},\ and\ \citenamefont
  {Schwenk}}]{Drischler:2014}%
  \BibitemOpen
  \bibfield  {author} {\bibinfo {author} {\bibfnamefont {C.}~\bibnamefont
  {Drischler}}, \bibinfo {author} {\bibfnamefont {V.}~\bibnamefont {Som\`a}}, \
  and\ \bibinfo {author} {\bibfnamefont {A.}~\bibnamefont {Schwenk}},\ }\href
  {\doibase 10.1103/PhysRevC.89.025806} {\bibfield  {journal} {\bibinfo
  {journal} {Phys. Rev. C}\ }\textbf {\bibinfo {volume} {89}},\ \bibinfo
  {pages} {025806} (\bibinfo {year} {2014})}\BibitemShut {NoStop}%
\bibitem [{\citenamefont {Frick}\ \emph {et~al.}(2005)\citenamefont {Frick},
  \citenamefont {M\"uther}, \citenamefont {Rios}, \citenamefont {Polls},\ and\
  \citenamefont {Ramos}}]{Frick:2005}%
  \BibitemOpen
  \bibfield  {author} {\bibinfo {author} {\bibfnamefont {T.}~\bibnamefont
  {Frick}}, \bibinfo {author} {\bibfnamefont {H.}~\bibnamefont {M\"uther}},
  \bibinfo {author} {\bibfnamefont {A.}~\bibnamefont {Rios}}, \bibinfo {author}
  {\bibfnamefont {A.}~\bibnamefont {Polls}}, \ and\ \bibinfo {author}
  {\bibfnamefont {A.}~\bibnamefont {Ramos}},\ }\href {\doibase
  10.1103/PhysRevC.71.014313} {\bibfield  {journal} {\bibinfo  {journal} {Phys.
  Rev. C}\ }\textbf {\bibinfo {volume} {71}},\ \bibinfo {pages} {014313}
  (\bibinfo {year} {2005})}\BibitemShut {NoStop}%
\bibitem [{\citenamefont {Bardeen}\ \emph {et~al.}(1957)\citenamefont
  {Bardeen}, \citenamefont {Cooper},\ and\ \citenamefont
  {Schrieffer}}]{Bardeen:1957}%
  \BibitemOpen
  \bibfield  {author} {\bibinfo {author} {\bibfnamefont {J.}~\bibnamefont
  {Bardeen}}, \bibinfo {author} {\bibfnamefont {L.~N.}\ \bibnamefont {Cooper}},
  \ and\ \bibinfo {author} {\bibfnamefont {J.~R.}\ \bibnamefont {Schrieffer}},\
  }\href {\doibase 10.1103/PhysRev.106.162} {\bibfield  {journal} {\bibinfo
  {journal} {Phys. Rev.}\ }\textbf {\bibinfo {volume} {106}},\ \bibinfo {pages}
  {162} (\bibinfo {year} {1957})}\BibitemShut {NoStop}%
\bibitem [{\citenamefont {Zhang}\ and\ \citenamefont
  {Krakauer}(2003)}]{Zhang:2003}%
  \BibitemOpen
  \bibfield  {author} {\bibinfo {author} {\bibfnamefont {S.}~\bibnamefont
  {Zhang}}\ and\ \bibinfo {author} {\bibfnamefont {H.}~\bibnamefont
  {Krakauer}},\ }\href {\doibase 10.1103/PhysRevLett.90.136401} {\bibfield
  {journal} {\bibinfo  {journal} {Phys. Rev. Lett.}\ }\textbf {\bibinfo
  {volume} {90}},\ \bibinfo {pages} {136401} (\bibinfo {year}
  {2003})}\BibitemShut {NoStop}%
\bibitem [{\citenamefont {Wiringa}\ and\ \citenamefont
  {Pieper}(2002)}]{Wiringa:2002}%
  \BibitemOpen
  \bibfield  {author} {\bibinfo {author} {\bibfnamefont {R.~B.}\ \bibnamefont
  {Wiringa}}\ and\ \bibinfo {author} {\bibfnamefont {S.~C.}\ \bibnamefont
  {Pieper}},\ }\href {\doibase 10.1103/PhysRevLett.89.182501} {\bibfield
  {journal} {\bibinfo  {journal} {Phys. Rev. Lett.}\ }\textbf {\bibinfo
  {volume} {89}},\ \bibinfo {pages} {182501} (\bibinfo {year}
  {2002})}\BibitemShut {NoStop}%
\bibitem [{sup()}]{supplemental}%
  \BibitemOpen
  \href@noop {} {}\bibinfo {note} {See Supplemental Material at [URL will be
  inserted by publisher]}\BibitemShut {NoStop}%
\bibitem [{\citenamefont {Sarsa}\ \emph {et~al.}(2003)\citenamefont {Sarsa},
  \citenamefont {Fantoni}, \citenamefont {Schmidt},\ and\ \citenamefont
  {Pederiva}}]{Sarsa:2003}%
  \BibitemOpen
  \bibfield  {author} {\bibinfo {author} {\bibfnamefont {A.}~\bibnamefont
  {Sarsa}}, \bibinfo {author} {\bibfnamefont {S.}~\bibnamefont {Fantoni}},
  \bibinfo {author} {\bibfnamefont {K.~E.}\ \bibnamefont {Schmidt}}, \ and\
  \bibinfo {author} {\bibfnamefont {F.}~\bibnamefont {Pederiva}},\ }\href
  {\doibase 10.1103/PhysRevC.68.024308} {\bibfield  {journal} {\bibinfo
  {journal} {Phys. Rev. C}\ }\textbf {\bibinfo {volume} {68}},\ \bibinfo
  {pages} {024308} (\bibinfo {year} {2003})}\BibitemShut {NoStop}%
\bibitem [{\citenamefont {{Gezerlis}}\ \emph {et~al.}()\citenamefont
  {{Gezerlis}}, \citenamefont {{Tews}}, \citenamefont {{Epelbaum}},
  \citenamefont {{Freunek}}, \citenamefont {{Gandolfi}}, \citenamefont
  {{Hebeler}}, \citenamefont {{Nogga}},\ and\ \citenamefont
  {{Schwenk}}}]{Gezerlis:2014}%
  \BibitemOpen
  \bibfield  {author} {\bibinfo {author} {\bibfnamefont {A.}~\bibnamefont
  {{Gezerlis}}}, \bibinfo {author} {\bibfnamefont {I.}~\bibnamefont {{Tews}}},
  \bibinfo {author} {\bibfnamefont {E.}~\bibnamefont {{Epelbaum}}}, \bibinfo
  {author} {\bibfnamefont {M.}~\bibnamefont {{Freunek}}}, \bibinfo {author}
  {\bibfnamefont {S.}~\bibnamefont {{Gandolfi}}}, \bibinfo {author}
  {\bibfnamefont {K.}~\bibnamefont {{Hebeler}}}, \bibinfo {author}
  {\bibfnamefont {A.}~\bibnamefont {{Nogga}}}, \ and\ \bibinfo {author}
  {\bibfnamefont {A.}~\bibnamefont {{Schwenk}}},\ }\href@noop {} {\bibinfo
  {journal} {arXiv:1406.0454}\ }\BibitemShut {NoStop}%
\bibitem [{\citenamefont {Pudliner}\ \emph {et~al.}(1997)\citenamefont
  {Pudliner}, \citenamefont {Pandharipande}, \citenamefont {Carlson},
  \citenamefont {Pieper},\ and\ \citenamefont {Wiringa}}]{Pudliner:1997}%
  \BibitemOpen
\bibfield  {journal} {  }\bibfield  {author} {\bibinfo {author} {\bibfnamefont
  {B.~S.}\ \bibnamefont {Pudliner}}, \bibinfo {author} {\bibfnamefont {V.~R.}\
  \bibnamefont {Pandharipande}}, \bibinfo {author} {\bibfnamefont
  {J.}~\bibnamefont {Carlson}}, \bibinfo {author} {\bibfnamefont {S.~C.}\
  \bibnamefont {Pieper}}, \ and\ \bibinfo {author} {\bibfnamefont {R.~B.}\
  \bibnamefont {Wiringa}},\ }\href {\doibase 10.1103/PhysRevC.56.1720}
  {\bibfield  {journal} {\bibinfo  {journal} {Phys. Rev. C}\ }\textbf {\bibinfo
  {volume} {56}},\ \bibinfo {pages} {1720} (\bibinfo {year}
  {1997})}\BibitemShut {NoStop}%
\bibitem [{\citenamefont {Lagaris}\ and\ \citenamefont
  {Pandharipande}(1981{\natexlab{b}})}]{Lagaris:1981}%
  \BibitemOpen
  \bibfield  {author} {\bibinfo {author} {\bibfnamefont {I.}~\bibnamefont
  {Lagaris}}\ and\ \bibinfo {author} {\bibfnamefont {V.}~\bibnamefont
  {Pandharipande}},\ }\href {\doibase
  http://dx.doi.org/10.1016/0375-9474(81)90241-4} {\bibfield  {journal}
  {\bibinfo  {journal} {Nuclear Physics A}\ }\textbf {\bibinfo {volume}
  {359}},\ \bibinfo {pages} {349 } (\bibinfo {year}
  {1981}{\natexlab{b}})}\BibitemShut {NoStop}%
\bibitem [{\citenamefont {Sorella}(2001)}]{Sorella:2001}%
  \BibitemOpen
  \bibfield  {author} {\bibinfo {author} {\bibfnamefont {S.}~\bibnamefont
  {Sorella}},\ }\href {\doibase 10.1103/PhysRevB.64.024512} {\bibfield
  {journal} {\bibinfo  {journal} {Phys. Rev. B}\ }\textbf {\bibinfo {volume}
  {64}},\ \bibinfo {pages} {024512} (\bibinfo {year} {2001})}\BibitemShut
  {NoStop}%
\bibitem [{\citenamefont {Maris}\ \emph
  {et~al.}(2013{\natexlab{b}})\citenamefont {Maris}, \citenamefont {Vary},
  \citenamefont {Gandolfi}, \citenamefont {Carlson},\ and\ \citenamefont
  {Pieper}}]{Maris:2013b}%
  \BibitemOpen
  \bibfield  {author} {\bibinfo {author} {\bibfnamefont {P.}~\bibnamefont
  {Maris}}, \bibinfo {author} {\bibfnamefont {J.~P.}\ \bibnamefont {Vary}},
  \bibinfo {author} {\bibfnamefont {S.}~\bibnamefont {Gandolfi}}, \bibinfo
  {author} {\bibfnamefont {J.}~\bibnamefont {Carlson}}, \ and\ \bibinfo
  {author} {\bibfnamefont {S.~C.}\ \bibnamefont {Pieper}},\ }\href {\doibase
  10.1103/PhysRevC.87.054318} {\bibfield  {journal} {\bibinfo  {journal} {Phys.
  Rev. C}\ }\textbf {\bibinfo {volume} {87}},\ \bibinfo {pages} {054318}
  (\bibinfo {year} {2013}{\natexlab{b}})}\BibitemShut {NoStop}%
\bibitem [{\citenamefont {Gandolfi}\ \emph {et~al.}(2008)\citenamefont
  {Gandolfi}, \citenamefont {Illarionov}, \citenamefont {Fantoni},
  \citenamefont {Pederiva},\ and\ \citenamefont {Schmidt}}]{Gandolfi:2008}%
  \BibitemOpen
  \bibfield  {author} {\bibinfo {author} {\bibfnamefont {S.}~\bibnamefont
  {Gandolfi}}, \bibinfo {author} {\bibfnamefont {A.~Y.}\ \bibnamefont
  {Illarionov}}, \bibinfo {author} {\bibfnamefont {S.}~\bibnamefont {Fantoni}},
  \bibinfo {author} {\bibfnamefont {F.}~\bibnamefont {Pederiva}}, \ and\
  \bibinfo {author} {\bibfnamefont {K.~E.}\ \bibnamefont {Schmidt}},\ }\href
  {\doibase 10.1103/PhysRevLett.101.132501} {\bibfield  {journal} {\bibinfo
  {journal} {Phys. Rev. Lett.}\ }\textbf {\bibinfo {volume} {101}},\ \bibinfo
  {pages} {132501} (\bibinfo {year} {2008})}\BibitemShut {NoStop}%
\bibitem [{\citenamefont {Gandolfi}\ \emph
  {et~al.}(2009{\natexlab{b}})\citenamefont {Gandolfi}, \citenamefont
  {Illarionov}, \citenamefont {Pederiva}, \citenamefont {Schmidt},\ and\
  \citenamefont {Fantoni}}]{Gandolfi:2009b}%
  \BibitemOpen
  \bibfield  {author} {\bibinfo {author} {\bibfnamefont {S.}~\bibnamefont
  {Gandolfi}}, \bibinfo {author} {\bibfnamefont {A.~Y.}\ \bibnamefont
  {Illarionov}}, \bibinfo {author} {\bibfnamefont {F.}~\bibnamefont
  {Pederiva}}, \bibinfo {author} {\bibfnamefont {K.~E.}\ \bibnamefont
  {Schmidt}}, \ and\ \bibinfo {author} {\bibfnamefont {S.}~\bibnamefont
  {Fantoni}},\ }\href {\doibase 10.1103/PhysRevC.80.045802} {\bibfield
  {journal} {\bibinfo  {journal} {Phys. Rev. C}\ }\textbf {\bibinfo {volume}
  {80}},\ \bibinfo {pages} {045802} (\bibinfo {year}
  {2009}{\natexlab{b}})}\BibitemShut {NoStop}%
\bibitem [{\citenamefont {Hagen}\ \emph
  {et~al.}(2012{\natexlab{b}})\citenamefont {Hagen}, \citenamefont
  {Hjorth-Jensen}, \citenamefont {Jansen}, \citenamefont {Machleidt},\ and\
  \citenamefont {Papenbrock}}]{Hagen:2012b}%
  \BibitemOpen
  \bibfield  {author} {\bibinfo {author} {\bibfnamefont {G.}~\bibnamefont
  {Hagen}}, \bibinfo {author} {\bibfnamefont {M.}~\bibnamefont
  {Hjorth-Jensen}}, \bibinfo {author} {\bibfnamefont {G.~R.}\ \bibnamefont
  {Jansen}}, \bibinfo {author} {\bibfnamefont {R.}~\bibnamefont {Machleidt}}, \
  and\ \bibinfo {author} {\bibfnamefont {T.}~\bibnamefont {Papenbrock}},\
  }\href {\doibase 10.1103/PhysRevLett.108.242501} {\bibfield  {journal}
  {\bibinfo  {journal} {Phys. Rev. Lett.}\ }\textbf {\bibinfo {volume} {108}},\
  \bibinfo {pages} {242501} (\bibinfo {year} {2012}{\natexlab{b}})}\BibitemShut
  {NoStop}%
\bibitem [{\citenamefont {Hagen}\ \emph
  {et~al.}(2012{\natexlab{c}})\citenamefont {Hagen}, \citenamefont
  {Hjorth-Jensen}, \citenamefont {Jansen}, \citenamefont {Machleidt},\ and\
  \citenamefont {Papenbrock}}]{Hagen:2012}%
  \BibitemOpen
  \bibfield  {author} {\bibinfo {author} {\bibfnamefont {G.}~\bibnamefont
  {Hagen}}, \bibinfo {author} {\bibfnamefont {M.}~\bibnamefont
  {Hjorth-Jensen}}, \bibinfo {author} {\bibfnamefont {G.~R.}\ \bibnamefont
  {Jansen}}, \bibinfo {author} {\bibfnamefont {R.}~\bibnamefont {Machleidt}}, \
  and\ \bibinfo {author} {\bibfnamefont {T.}~\bibnamefont {Papenbrock}},\
  }\href {\doibase 10.1103/PhysRevLett.109.032502} {\bibfield  {journal}
  {\bibinfo  {journal} {Phys. Rev. Lett.}\ }\textbf {\bibinfo {volume} {109}},\
  \bibinfo {pages} {032502} (\bibinfo {year} {2012}{\natexlab{c}})}\BibitemShut
  {NoStop}%
\bibitem [{\citenamefont {Hergert}\ \emph {et~al.}(2013)\citenamefont
  {Hergert}, \citenamefont {Bogner}, \citenamefont {Binder}, \citenamefont
  {Calci}, \citenamefont {Langhammer}, \citenamefont {Roth},\ and\
  \citenamefont {Schwenk}}]{Hergert:2013}%
  \BibitemOpen
  \bibfield  {author} {\bibinfo {author} {\bibfnamefont {H.}~\bibnamefont
  {Hergert}}, \bibinfo {author} {\bibfnamefont {S.~K.}\ \bibnamefont {Bogner}},
  \bibinfo {author} {\bibfnamefont {S.}~\bibnamefont {Binder}}, \bibinfo
  {author} {\bibfnamefont {A.}~\bibnamefont {Calci}}, \bibinfo {author}
  {\bibfnamefont {J.}~\bibnamefont {Langhammer}}, \bibinfo {author}
  {\bibfnamefont {R.}~\bibnamefont {Roth}}, \ and\ \bibinfo {author}
  {\bibfnamefont {A.}~\bibnamefont {Schwenk}},\ }\href {\doibase
  10.1103/PhysRevC.87.034307} {\bibfield  {journal} {\bibinfo  {journal} {Phys.
  Rev. C}\ }\textbf {\bibinfo {volume} {87}},\ \bibinfo {pages} {034307}
  (\bibinfo {year} {2013})}\BibitemShut {NoStop}%
\bibitem [{\citenamefont {Cipollone}\ \emph {et~al.}(2013)\citenamefont
  {Cipollone}, \citenamefont {Barbieri},\ and\ \citenamefont
  {Navr\'atil}}]{Cipollone:2013}%
  \BibitemOpen
  \bibfield  {author} {\bibinfo {author} {\bibfnamefont {A.}~\bibnamefont
  {Cipollone}}, \bibinfo {author} {\bibfnamefont {C.}~\bibnamefont {Barbieri}},
  \ and\ \bibinfo {author} {\bibfnamefont {P.}~\bibnamefont {Navr\'atil}},\
  }\href {\doibase 10.1103/PhysRevLett.111.062501} {\bibfield  {journal}
  {\bibinfo  {journal} {Phys. Rev. Lett.}\ }\textbf {\bibinfo {volume} {111}},\
  \bibinfo {pages} {062501} (\bibinfo {year} {2013})}\BibitemShut {NoStop}%
\bibitem [{\citenamefont {Ekstr\"om}\ \emph {et~al.}(2013)\citenamefont
  {Ekstr\"om}, \citenamefont {Baardsen}, \citenamefont {Forss\'en},
  \citenamefont {Hagen}, \citenamefont {Hjorth-Jensen}, \citenamefont {Jansen},
  \citenamefont {Machleidt}, \citenamefont {Nazarewicz}, \citenamefont
  {Papenbrock}, \citenamefont {Sarich},\ and\ \citenamefont
  {Wild}}]{Ekstrom:2013}%
  \BibitemOpen
  \bibfield  {author} {\bibinfo {author} {\bibfnamefont {A.}~\bibnamefont
  {Ekstr\"om}}, \bibinfo {author} {\bibfnamefont {G.}~\bibnamefont {Baardsen}},
  \bibinfo {author} {\bibfnamefont {C.}~\bibnamefont {Forss\'en}}, \bibinfo
  {author} {\bibfnamefont {G.}~\bibnamefont {Hagen}}, \bibinfo {author}
  {\bibfnamefont {M.}~\bibnamefont {Hjorth-Jensen}}, \bibinfo {author}
  {\bibfnamefont {G.~R.}\ \bibnamefont {Jansen}}, \bibinfo {author}
  {\bibfnamefont {R.}~\bibnamefont {Machleidt}}, \bibinfo {author}
  {\bibfnamefont {W.}~\bibnamefont {Nazarewicz}}, \bibinfo {author}
  {\bibfnamefont {T.}~\bibnamefont {Papenbrock}}, \bibinfo {author}
  {\bibfnamefont {J.}~\bibnamefont {Sarich}}, \ and\ \bibinfo {author}
  {\bibfnamefont {S.~M.}\ \bibnamefont {Wild}},\ }\href {\doibase
  10.1103/PhysRevLett.110.192502} {\bibfield  {journal} {\bibinfo  {journal}
  {Phys. Rev. Lett.}\ }\textbf {\bibinfo {volume} {110}},\ \bibinfo {pages}
  {192502} (\bibinfo {year} {2013})}\BibitemShut {NoStop}%
\bibitem [{\citenamefont {Schmidt}\ \emph {et~al.}(1981)\citenamefont
  {Schmidt}, \citenamefont {Lee}, \citenamefont {Kalos},\ and\ \citenamefont
  {Chester}}]{Schmidt:1981}%
  \BibitemOpen
  \bibfield  {author} {\bibinfo {author} {\bibfnamefont {K.~E.}\ \bibnamefont
  {Schmidt}}, \bibinfo {author} {\bibfnamefont {M.~A.}\ \bibnamefont {Lee}},
  \bibinfo {author} {\bibfnamefont {M.~H.}\ \bibnamefont {Kalos}}, \ and\
  \bibinfo {author} {\bibfnamefont {G.~V.}\ \bibnamefont {Chester}},\ }\href
  {\doibase 10.1103/PhysRevLett.47.807} {\bibfield  {journal} {\bibinfo
  {journal} {Phys. Rev. Lett.}\ }\textbf {\bibinfo {volume} {47}},\ \bibinfo
  {pages} {807} (\bibinfo {year} {1981})}\BibitemShut {NoStop}%
\bibitem [{\citenamefont {Lee}\ \emph {et~al.}(1981)\citenamefont {Lee},
  \citenamefont {Schmidt}, \citenamefont {Kalos},\ and\ \citenamefont
  {Chester}}]{Lee:1981}%
  \BibitemOpen
  \bibfield  {author} {\bibinfo {author} {\bibfnamefont {M.~A.}\ \bibnamefont
  {Lee}}, \bibinfo {author} {\bibfnamefont {K.~E.}\ \bibnamefont {Schmidt}},
  \bibinfo {author} {\bibfnamefont {M.~H.}\ \bibnamefont {Kalos}}, \ and\
  \bibinfo {author} {\bibfnamefont {G.~V.}\ \bibnamefont {Chester}},\ }\href
  {\doibase 10.1103/PhysRevLett.46.728} {\bibfield  {journal} {\bibinfo
  {journal} {Phys. Rev. Lett.}\ }\textbf {\bibinfo {volume} {46}},\ \bibinfo
  {pages} {728} (\bibinfo {year} {1981})}\BibitemShut {NoStop}%
\bibitem [{\citenamefont {Holzmann}\ \emph {et~al.}(2003)\citenamefont
  {Holzmann}, \citenamefont {Ceperley}, \citenamefont {Pierleoni},\ and\
  \citenamefont {Esler}}]{Holtzmann:2003}%
  \BibitemOpen
  \bibfield  {author} {\bibinfo {author} {\bibfnamefont {M.}~\bibnamefont
  {Holzmann}}, \bibinfo {author} {\bibfnamefont {D.~M.}\ \bibnamefont
  {Ceperley}}, \bibinfo {author} {\bibfnamefont {C.}~\bibnamefont {Pierleoni}},
  \ and\ \bibinfo {author} {\bibfnamefont {K.}~\bibnamefont {Esler}},\ }\href
  {\doibase 10.1103/PhysRevE.68.046707} {\bibfield  {journal} {\bibinfo
  {journal} {Phys. Rev. E}\ }\textbf {\bibinfo {volume} {68}},\ \bibinfo
  {pages} {046707} (\bibinfo {year} {2003})}\BibitemShut {NoStop}%
\bibitem [{\citenamefont {Tanatar}\ and\ \citenamefont
  {Ceperley}(1989)}]{Tanatar:1989}%
  \BibitemOpen
  \bibfield  {author} {\bibinfo {author} {\bibfnamefont {B.}~\bibnamefont
  {Tanatar}}\ and\ \bibinfo {author} {\bibfnamefont {D.~M.}\ \bibnamefont
  {Ceperley}},\ }\href {\doibase 10.1103/PhysRevB.39.5005} {\bibfield
  {journal} {\bibinfo  {journal} {Phys. Rev. B}\ }\textbf {\bibinfo {volume}
  {39}},\ \bibinfo {pages} {5005} (\bibinfo {year} {1989})}\BibitemShut
  {NoStop}%
\bibitem [{\citenamefont {Pilati}()}]{Pilati:private}%
  \BibitemOpen
  \bibfield  {author} {\bibinfo {author} {\bibfnamefont {S.}~\bibnamefont
  {Pilati}},\ }\href@noop {} {\ }\bibinfo {note} {Private
  communication}\BibitemShut {NoStop}%
\bibitem [{\citenamefont {Pilati}\ \emph {et~al.}(2010)\citenamefont {Pilati},
  \citenamefont {Bertaina}, \citenamefont {Giorgini},\ and\ \citenamefont
  {Troyer}}]{Pilati:2010}%
  \BibitemOpen
  \bibfield  {author} {\bibinfo {author} {\bibfnamefont {S.}~\bibnamefont
  {Pilati}}, \bibinfo {author} {\bibfnamefont {G.}~\bibnamefont {Bertaina}},
  \bibinfo {author} {\bibfnamefont {S.}~\bibnamefont {Giorgini}}, \ and\
  \bibinfo {author} {\bibfnamefont {M.}~\bibnamefont {Troyer}},\ }\href
  {\doibase 10.1103/PhysRevLett.105.030405} {\bibfield  {journal} {\bibinfo
  {journal} {Phys. Rev. Lett.}\ }\textbf {\bibinfo {volume} {105}},\ \bibinfo
  {pages} {030405} (\bibinfo {year} {2010})}\BibitemShut {NoStop}%
\bibitem [{\citenamefont {Lovato}\ \emph {et~al.}(2011)\citenamefont {Lovato},
  \citenamefont {Benhar}, \citenamefont {Fantoni}, \citenamefont {Illarionov},\
  and\ \citenamefont {Schmidt}}]{Lovato:2011}%
  \BibitemOpen
  \bibfield  {author} {\bibinfo {author} {\bibfnamefont {A.}~\bibnamefont
  {Lovato}}, \bibinfo {author} {\bibfnamefont {O.}~\bibnamefont {Benhar}},
  \bibinfo {author} {\bibfnamefont {S.}~\bibnamefont {Fantoni}}, \bibinfo
  {author} {\bibfnamefont {A.~Y.}\ \bibnamefont {Illarionov}}, \ and\ \bibinfo
  {author} {\bibfnamefont {K.~E.}\ \bibnamefont {Schmidt}},\ }\href {\doibase
  10.1103/PhysRevC.83.054003} {\bibfield  {journal} {\bibinfo  {journal} {Phys.
  Rev. C}\ }\textbf {\bibinfo {volume} {83}},\ \bibinfo {pages} {054003}
  (\bibinfo {year} {2011})}\BibitemShut {NoStop}%
\end{thebibliography}

%

\end{document}